# Properties of Nanocrystalline Silicon Probed by Optomechanics

*Daniel Navarro-Urrios[1,2*], Martín F. Colombano[1,2], Jeremie Maire[1], Emigdio Chávez-Ángel[1], Guillermo Arregui[1,3], Néstor E. Capuj[4,5], Arnaud Devos[6], Amadeu Griol[7], Laurent Bellieres[7], Alejandro Martínez[7], Kestutis Grigoras[8], Teija Häkkinen[8], Jaakko Saarilahti[8], Tapani Makkonen[8], Clivia M. Sotomayor-Torres[1,9], Jouni Ahopelto[8]*

[1] Catalan Institute of Nanoscience and Nanotechnology (ICN2), CSIC and BIST, Campus UAB, Bellaterra, 08193 Barcelona, Spain

[2] MIND-IN2UB, Departament d'Enginyeria Electrònica i Biomèdica, Facultat de Física, Universitat de Barcelona, Martí i Franquès 1, 08028 Barcelona, Spain

[3] Depto. Física, Universidad Autónoma de Barcelona, Bellaterra, 08193 Barcelona, Spain

[4] Depto. Física, Universidad de La Laguna, 38200 San Cristóbal de La Laguna, Spain

[5] Instituto Universitario de Materiales y Nanotecnología, Universidad de La Laguna, 38071 Santa Cruz de Tenerife, Spain

[6] Institut d'Electronique, de Microélectronique et de Nanotechnologie, Unité Mixte de Recherche, CNRS 8250, Avenue Poincaré, CS 60069, F-59652 Villeneuve d'Ascq Cedex, France

[7] Nanophotonics Technology Center, Universitat Politècnica de València, 46022 València, Spain

[8] VTT Technical Research Centre of Finland Ltd, P.O. Box 1000, FI-02044 VTT, Espoo, Finland

[9] Catalan Institute for Research and Advances Studies ICREA, 08010 Barcelona, Spain

KEYWORDS: nanocrystalline silicon, cavity optomechanics, annealing.

ABSTRACT

Nanocrystalline materials exhibit properties that can differ substantially from those of their single crystal counterparts. As such, they provide ways to enhance and optimise their functionality for devices and applications. Here we report on the optical, mechanical and thermal properties of nanocrystalline silicon probed by means of optomechanical nanobeams to extract information of the dynamics of optical absorption, mechanical losses, heat generation and dissipation. The optomechanical nanobeams are fabricated using nanocrystalline films prepared by annealing amorphous silicon layers at different temperatures. The resulting crystallite sizes and the stress in the films can be controlled by the annealing temperature and time and, consequently, the properties of the films can be tuned relatively freely, as demonstrated here by means of electron microscopy and Raman scattering. We show that the nanocrystallite size and the volume fraction of the grain boundaries play a key role in the dissipation rates through non-linear optical and thermal processes. Promising optical (13000) and mechanical (1700) quality factors were found in the optomechanical cavity realised in the nanocrystalline Si resulting from annealing at 950 ºC. The enhanced absorption and recombination rates via the intra-gap states and the reduced thermal conductivity boost the potential to exploit these non-linear effects in applications, including NEMS, phonon lasing and chaos-based devices.

1. **Introduction**

Nanoscale optomechanical cavities (OMC) [1] promise a route towards flexible, efficient and reliable interfaces to transfer classical and or quantum information between microwave and optical frequency domains in a single chip. In the non-linear regime these devices are very attractive for room temperature applications, such as mass sensing, [2] non-volatile memories, [3] chaos-based applications [4] and coupled oscillator networks for neuromorphic computing, [5], [6] among others. The strong interaction between light and mechanical modes has been demonstrated in nanoscale optomechanical cavities made of silicon nitride ($Si_3N_4$), [7] gallium arsenide (GaAs), [8] aluminum nitride (AlN), [9], [10] diamond, [11], [12] and crystalline silicon (c-Si). [13], [14] In particular, the well-established silicon-on-insulator (SOI) technology has been the cornerstone in most of the leading experimental results, including ground state cooling [15] and the observation of single photon/phonon quantum correlations, [16] to name but a few. An interesting alternative to SOI is nanocrystalline silicon (nc-Si)-on-insulator (nc-SOI). Nanocrystalline silicon is formed by thermal annealing of originally amorphous silicon to transform it into a polycrystalline material. The process was developed in the early 80's [17] and is used in MEMS technology due to the relatively easy tuning of mechanical, optical, electrical and thermal properties by doping, tailoring the size of the crystallites and the stress in the films. [17], [18] Recently, it was demonstrated that nc-Si is also an excellent and cost-competitive alternative to c-Si for optomechanical devices taking advantage of non-linear effects while operating in ambient conditions. [19] Non-linear dynamic functions, such as phonon lasing and chaos, were demonstrated with broader bandwidth than measured in equivalent SOI devices as well as wafers. [19] The dynamic behavior depends on the optical and mechanical losses and on heat generation and extraction rates of the OMCs. [4], [20] These properties can be tuned rather flexibly in nanocrystalline silicon films by varying the processing parameters. In this work, we use optomechanical nanobeams to probe the properties of nanocrystalline films, focusing on the optical and mechanical behavior and, especially, on dissipation rates. A set of nominally identical nanobeam OMCs on nc-SOI wafers annealed at different temperatures, resulting in films with different nanocrystallite size distribution and tensile stress were fabricated. The experimental results are compared with those obtained from geometrically identical OMCs fabricated on SOI wafers.

2. **Nanocrystalline silicon-on-insulator substrates**

The substrates for the optomechanical nanobeam cavities consisted of an un-doped, nominally 220 nm thick nc-Si film on a 1000 nm thick thermal oxide layer grown on low-doped 150 mm (100) Si wafers. The thick $SiO_2$ layer was grown by wet oxidation at 1050 °C. The measured resulting thickness was 1013 nm. On the oxide, a layer of amorphous Si was deposited at 574 °C from silane by low pressure chemical vapor deposition (LPCVD). The amorphous Si layer was converted to a nanocrystalline one by annealing it for 60 min at different temperatures $T_a$ to prepare the set of nc-SOI wafers for the OMCs. The thickness of the nc-Si films before and after annealing was measured using monitor wafers processed in the same batch as the device wafers. The thickness of the amorphous Si layer before annealing was 219±4 nm, measured by spectroscopic reflectometry at the center and in the middle of each quadrant on monitor wafers. After annealing at 750 and 950 °C, the measured thickness of both monitor wafers was 211±4 nm. The change in volume, when the material is transformed from amorphous to nanocrystalline, converts the originally compressive stress in the amorphous Si film to tensile stress in the nanocrystalline film. The amount of stress can be controlled by the annealing temperature and time. The measured tensile stress of the samples after annealing ranged from 90 MPa to 290 MPa. The residual stress was estimated from the change in the curvature of the wafers after the deposition and annealing of the nc-Si films using Stoney's equation [21] with the biaxial modulus for (100) Si wafers. [22] Table I summarizes the properties of the samples discussed in this work.

**Table I.** Summary of the main characteristics of the samples studied. The values measured on the wafers and those obtained from the OMCs are in the blue- and red-headed columns, respectively.

| Sample name | Description | $T_a$ (°C) | Thickness (nm) (monitor wafers) | Stress (MPa) | Refractive index @ 633 nm | Average nc size (nm) | Longitudinal sound velocity (m/s) | Intrinsic optical decay rate @ 1.55 µm ($\times 10^{11}$ s$^{-1}$) | Mechanical Q-factor @ 2.4 GHz | Temperature increase @ 2700 intracavity photons (K) | Ref. |
|---|---|---|---|---|---|---|---|---|---|---|---|
| **c-Si** | Reference Si | - | 220 | -39 | 3.88 | - | 8100 | 0.63 | 745 | 0.2 | [23], [24], this work |
| **a-Si** | Amorphous Si-reference | - | 219 | -190 | 4.36 | - | 7950 | - | - | - | This work |
| **OMS1** | Annealed a-Si | 650 | - | 290 | - | 163 | 8510 | 1.8 | 202 | 61 | This work |
| **OMS2** | Annealed a-Si | 750 | 211 | 250 | 3.93 | 171 | 8510 | 1.76 | 389 | - | This work |
| **OMS3** | Annealed a-Si | 850 | - | 170 | - | 187 | 8510 | 1.45 | 601 | 41 | This work |
| **OMS4** | Annealed a-Si | 950 | 211 | 90 | 3.93 (3.49@1550 nm) | 215 | 8510 | 1.14 | 1680 | 36 | This work |

The size of the nanocrystallites in the annealed films ranges from a few nanometers to roughly the thickness of the film, with the size distribution depending on the annealing temperature and time. To investigate the effect of annealing temperature on the microscopic structure of the films, a detailed study of the structure and size distribution of the nanocrystallites was carried out using transmission electron microscopy (TEM). A bright-field TEM micrograph of the sample OMS4 is shown in Figure 1A. The structure in the image is representative of the whole set of nc-Si samples. Selective area electron diffraction (SAED) confirmed that all the annealed samples are polycrystalline with no preferential crystalline orientation, see the inset to Figure 1A, in contrast to films deposited at higher temperature for which the initial growth mode is polycrystalline. The size distribution of the nanocrystallites was obtained by dark-field TEM analysis method (DF-TEM), as detailed in the Supporting Information. Briefly, an aperture is placed in the diffraction plane, effectively blocking electrons diffracted outside the range determined by the aperture. This creates images in which only nanocrystallites with a specific crystalline orientation appear bright on dark background. By moving the aperture, all orientations can be inspected. The size of the nanocrystallites is estimated using a threshold in the brightness of their images along two perpendicular directions fixed for each of the samples. The statistical analysis includes more than 500 nanocrystallites in each sample. The average size and size distribution were extracted from this data. The average size was found to increase with annealing temperature from 163 (annealing at 650 ºC) to 215 nm (annealing at 950 ºC) and the size distribution became broader the larger the nanocrystallites, as shown in Figure 1B. Thus, it is deduced that the volume fraction of the grain boundaries within the layers, representing the volume occupied by the nanocrystallite boundaries with respect to the total volume of the nc-Si layer ($\eta_{GB}$), decreases upon annealing.

We also investigated the nc-Si films by Raman spectroscopy using a 532 nm laser, keeping the laser power low enough to avoid heating effects, and compared the results with the reference c-Si samples measured spectra. The typical optical phonon mode of crystalline Si centered at 520 cm$^{-1}$ is observed in all samples and in the nc-Si it exhibits an asymmetric broadening in the low frequency side increasing the full width at half maximum (FWHM) of the peak. This broadening is typically associated to the presence of an initial amorphous phase, [25] nanocrystallites of different sizes [26] and to grain boundaries. [27] In this study the latter two are more likely since the amorphous phase contribution usually shows spectral features at significantly lower energies, around ~480 cm$^{-1}$. [28] This is further supported by the SAED images, which indicate that the films are polycrystalline with no halos arising from the amorphous phase. The FWHM of the peak decreases with $T_a$ (see the inset to Figure 1C), which is interpreted as arising from the smaller grain boundary fraction volume ($\eta_{GB}$ decreases) and the concomitant shift of the distribution towards larger nanocrystallites.

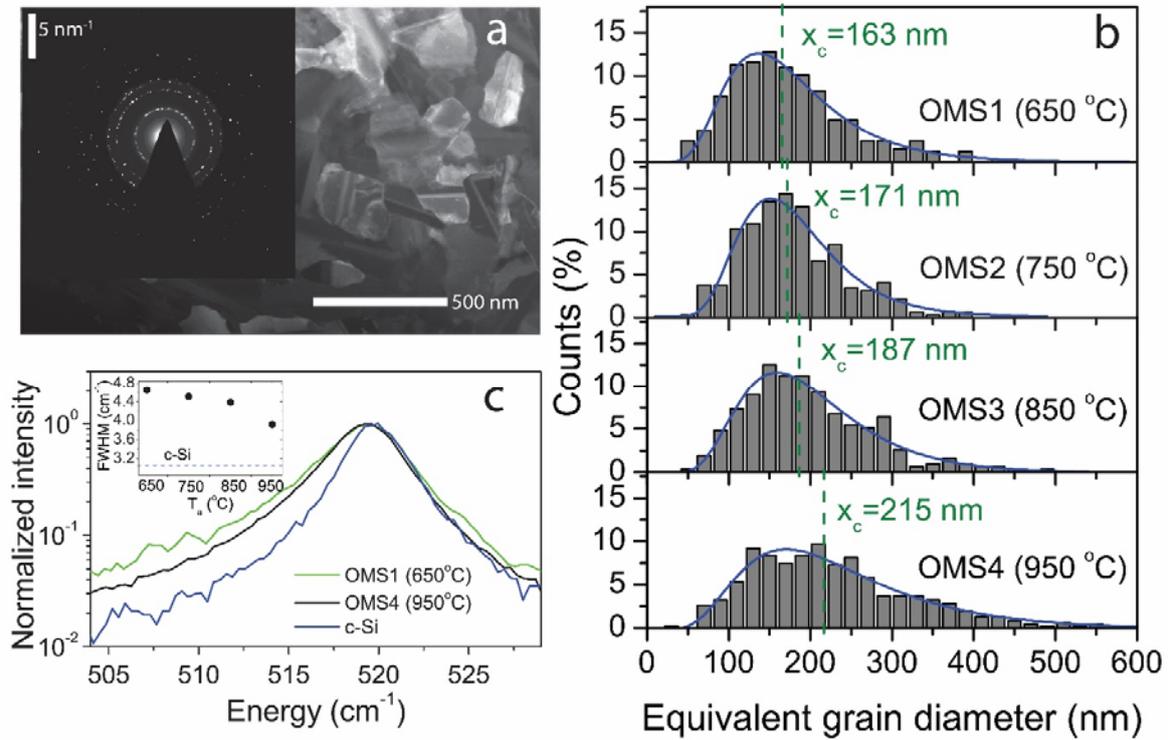

**Figure 1. Nanocrystallite size distribution in the nc-Si films.** (A) Bright field TEM image of the sample OMS4, showing the randomly oriented nanocrystallites. The inset shows a selective area electron diffraction (SAED) image of the sample. (B) Histograms of the crystallite size distribution of samples annealed at different temperatures $T_a$. The analysis includes more than 500 nanocrystallites in each sample and a log-normal distribution is used to fit the histograms. The average nanocrystallite size ($x_c$) is shown by the vertical dashed black lines. (C) Raman scattering spectra of the reference c-Si (blue) and the nc-Si layers annealed at 650 and 950ºC (green and black, respectively). The full-width-at-half-maximum of the Raman peak as a function of the annealing temperature is shown in the inset. The dashed horizontal line corresponds to the value measured from the c-Si reference sample.

The sound velocity of the nc-Si films was measured by picosecond acoustics. [29] Interestingly, the sound velocity in all annealed samples was 8510 m/s, which is slightly higher than that in (100) Si but lower than in (111) and (110) Si. This may be a consequence of the polycrystalline nature, averaging the velocities in different crystallographic directions. The sound velocity of the as-deposited amorphous Si was measured to be 7950 m/s, suggesting that the growth mode of the LPCVD Si at 574 ˚C is close to switching from amorphous to polycrystalline. This can also be seen in the SAED images of the as-deposited Si (see Supplementary Information).

### 3. Optomechanical characterisation of the nanobeams

Nanobeams with optomechanical cavities were fabricated in nc-SOI wafers to study the coupling of light and mechanical vibrations in nanocrystalline silicon. The fabrication process of the OMCs is described in detail in Ref. [30]. Briefly, the OMC geometry is based on a unit-cell consisting of a parallelogram with a cylindrical hole in the center and two symmetric stubs on the sides as shown in the SEM image in Figure 2. The cavity region in the center of the beam consists of 12 cells with the pitch ($a$), the radius of the holes ($r$) and the length of the stubs ($d$) decreasing quadratically towards the center of the beam. Ten-period mirrors are placed on both sides of the central region. The nominal geometrical values of the cells of the mirror are $a = 500$ nm, $r = 150$ nm, and $d = 250$ nm. The ratio of the geometrical parameters of the central cell with respect to those of the mirror cells is 0.85. The length of the OMC beam is 15 μm. After patterning, the beam is released by removing the oxide layer beneath the beam.

The characterization of the optical and mechanical modes supported by the OMCs was performed using the set-up shown in Figure 2. The emission of two tunable lasers covering the spectral around 1.55 μm (L1 and L2) with independently controlled polarization states are multiplexed into a tapered fiber. The tapered part of the fiber has a loop to enhance the spatial resolution and help aligning it with the beam. The loop is placed parallel to the OMC, in contact to an edge of the released window with a gap between the fiber and the OMC of about 200 nm. The long tail of the evanescent field of the taper excites locally the resonant optical modes of the OMC. The optical signal is measured either in transmission or in reflection using fast InGaAs photoreceivers PD1 and PD2 with bandwidths of 12 GHz. Once in optical resonance the mechanical motion of the cavity, activated by the thermal Langevin force, modulates the cavity mode and the signal is recorded using a spectrum analyzer (SA) with a bandwidth of 13.5 GHz. All the measurements are performed in an anti-vibration cage at atmospheric pressure and temperature.

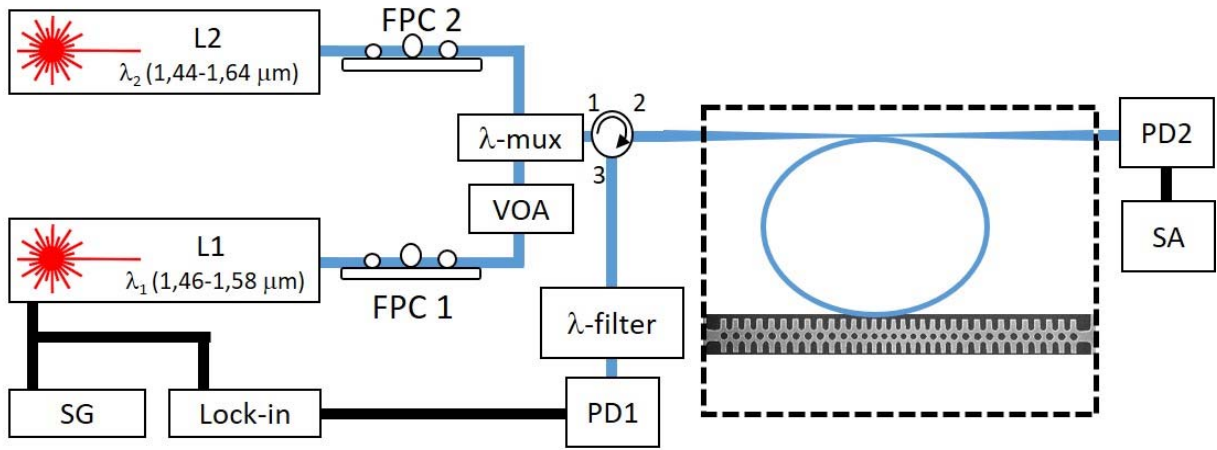

**Figure 2. Schematics of the experimental setup to measure the properties of the optomechanical nanobeams**. A tapered fiber loop is placed in close proximity to the OMC nanobeam, shown in the SEM image (not in scale). L1 and L2 are tunable lasers, λ-mux is a wavelength multiplexer, λ-filter is a Fabry-Perot wavelength filter, SG denotes a signal generator, VOA a variable optical attenuator and FPC1 and FPC2 are the fiber polarizer controllers, SA stands for spectrum analyser and PD1 and PD2 denote the photodetectors. The signals can be detected in transmission or reflection mode.

a. **Mechanical properties**

When the excitation laser wavelength is in resonance with the optical cavity mode, it is possible to transduce thermally activated mechanical modes in the sample. RF signals of the mechanical breathing modes appearing at about 2.4 GHz are shown in Figure 3 in a normalized frequency scale. We studied the behavior of the mechanical quality factor $Q_m$ and its dependence on $T_a$. The results are shown in the inset to Figure 3. It is clear that $Q_m$ increases with $T_a$, whereas the quality factor decreases with increasing tensile stress in the nc-Si layer, as opposed to c-Si or $Si_3N_4$ based nanomechanical resonators. [31] For the highest annealing temperature, $Q_m$ is about $1.7 \times 10^3$, which is a factor of two to three times higher than what was found in equivalent c-Si OMCs at room temperature.

Thermoelastic damping (TED), resulting from heat generation due to vibration and dissipation from thermal diffusion, is often the dominant factor determining the mechanical energy dissipation in micro and nanosized beam resonators. [32] An experimentally validated theory to describe the TED was first established by Zener for homogeneous materials. [33] This theory was recently extended to polycrystalline materials, in which an enhancement of the TED is expected due to both the reduction of the overall thermal diffusivity of the material and to the inter-crystalline damping as a consequence of additional dilatational strains created at the grain boundaries. [34] It is also well known that providing mechanical tensile pre-stress, or strain, is an effective way to reduce mechanical losses, [31], [35] a step in part associated to mitigation of the TED. [36]

The results shown in Figure 3 reflect that a balance between the two contributions described above determine the overall mechanical dissipation rates in the mechanical modes under study: (i) mechanical tensile strain and (ii) thermoelastic damping. The results suggest that TED has stronger impact than the tensile strain on the $Q_m$ of the nc-Si OMC. Interestingly, the nanobeam OMS4 fabricated in the wafer annealed at the highest $T_a$ has a higher $Q_m$ than the equivalent c-Si beam. It is worth noting that the small compressive stress in the c-SOI sample leads to buckling of the released OMC, which may affect the $Q_m$.

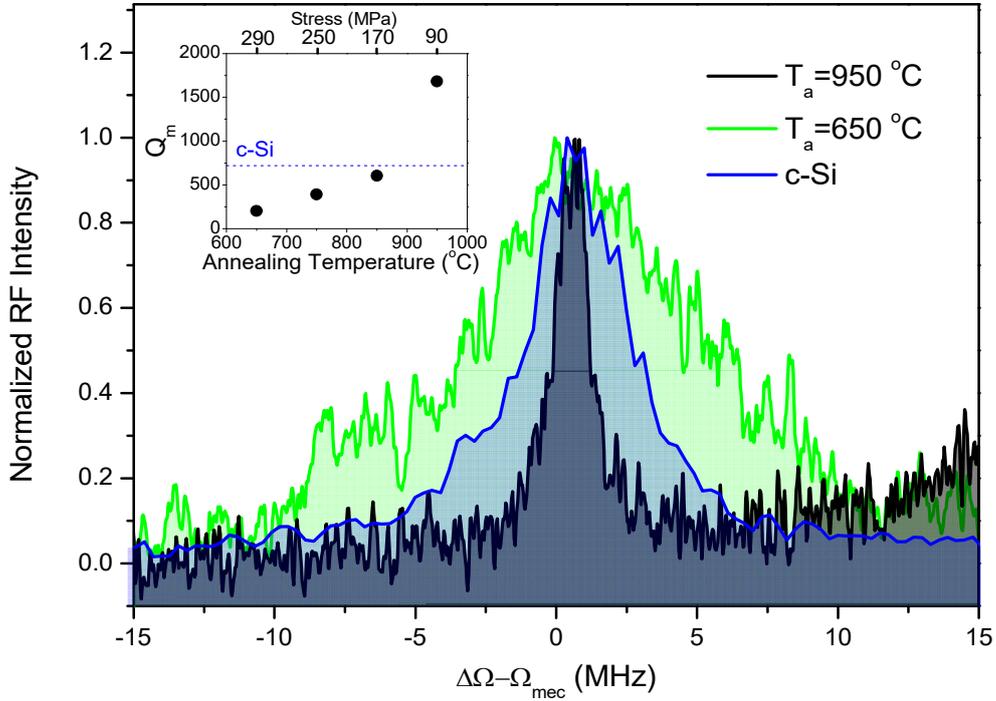

**Figure 3. Normalized RF spectra of a mechanical breathing mode in optomechanical nanobeams.** The mode appearing at about 2.4 GHz was measured in the nc-Si OMC1 and OMS2 and the reference c-Si OMC. The frequency scale is normalized to allow comparison of the linewidths of different samples. The inset shows the evolution of the mechanical Q-factor as a function of annealing temperature and the measured tensile stress. The dashed horizontal line corresponds to the value measured in the c-Si reference OMC.

### b. Optomechanical and thermal properties

Low excitation power optical spectra measured in the OMCs are shown in Figure 4A. It is worth noting that from now on we deal with the fundamental and the 2$^{nd}$ order optical modes of the OMCs (indicated in the notation with subindices 1 and 2, respectively). The obtained spectra confirmed that the frequencies of the resonant peaks are similar in all samples within a range of a few nanometers. This is not surprising, since the nc-Si layers in all samples have the same thickness and refractive index as measured by spectroscopic reflectometry (see Table I). However, the intrinsic optical decay rate of the fundamental mode $\kappa_{i,1}$ at the resonant wavelength $\lambda_{r,1} \approx 1.54$ μm, extracted indirectly by measuring the overall decay rate ($\kappa_1$) and the coupled power fraction in resonance, [19] decreases in samples annealed at higher temperatures, as shown in Figure 4B. The lowest $\kappa_{i,1}$ value is still a factor of 1.5 higher than that of a c-Si OMC with the same geometry, and leads to an intrinsic optical Q-factor of $Q_{i,1}=1.3 \times 10^4$ at RT. We deduce that the material losses are the dominant loss mechanisms in nc-Si based OMCs, since the radiative losses depend on the geometrical shape and thus can be assumed to be independent from $T_a$. We postulate that the main intrinsic optical loss mechanism arises from scattering at the grain boundaries and from the carrier involving interface states within the band gap. As the wavelength is below the band gap energy of silicon, the carrier generation taking place via gap states and the ensuing relaxation with phonon emission, i.e., in a Shockley-Read-Hall recombination mechanism, [37] determine the optical and thermal properties of nc-Si OMCs. The optical losses become smaller in samples annealed with increasing $T_a$, explained above by the drop in grain boundary volume and, possibly, of the trap density at the boundaries.

Since at optical energies falling below the band gap Si-based optical resonators have three main power-dependent optical loss mechanisms: i) two-photon absorption (TPA), ii) trap-assisted carrier generation, and iii) free-carrier absorption within the conduction and valence bands, the dependence of $\kappa_{i,1}$ on the number of intra-cavity photons $n_o$ was studied. The first scales with $n_o^2$ while the second is linear in $n_o$ and depends on the trap density. The third mechanism is linear with the excited carrier density $N$ and the number of photons in the cavity $n_o$.

In what follows carrier absorption (CA), consisting of the latter two mechanisms, is discussed as giving rise to the main optical loss mechanism in nc-Si. The trap states inside the band gap are thermally populated up to the Fermi level lying inside the band gap in our samples. They provide a reservoir for carrier generation to the conduction band and the empty trap states then generate holes. To quantify the roles of these optical loss mechanisms, a pump-and-probe technique was used in which the 2$^{nd}$ order mode of the OMC is excited using the laser L2 with input power ($P_{in}$) of the order of hundreds of µW, while the fundamental mode is probed with the laser L1 with sufficiently low power so that its effect on $\kappa_{i,1}$ can be neglected, see Figure 4A. Given that the number of intra-cavity photons $n_o$ follows the Lorentzian shape of an optical resonance, i.e., $n_o = n_{o,max} \Delta\lambda_{r,2}^2 / 4(\lambda_{L2} - \lambda_{r,2})^2 + \Delta\lambda_{r,2}^2$, the followed procedure to adjust $n_o$ is to fix the input power and decrease the relative detuning between the laser L2 wavelength $\lambda_{L2}$ and the resonance wavelength of the 2$^{nd}$ order mode $\lambda_{r,2}$, When the laser is in resonance with the 2$^{nd}$ order mode, the expression becomes $n_o = n_{o,max} = 2P_{in}\lambda_{r,2}\kappa_{e,2} / \kappa_2^2 hc$, $\kappa_{e,2}$ and $\kappa_2$ being the extrinsic and overall optical decay rates of the 2$^{nd}$ order mode, respectively, $h$ is the Planck's constant and $c$ the speed of light. In this particular experiment, the light extracted from the fundamental cavity mode is collected in a reflection configuration using a circulator and a tunable Fabry-Perot filter in resonance with the laser L1 to filter out the contribution of light from the 2$^{nd}$ mode. The output of the laser L1 is modulated using a signal generator, the output of which is used as the reference to a lock-in amplifier receiving the signal from the photodiode PD2 to improve the signal-to-noise ratio, see Figure 2. We verified experimentally that $\kappa_2$ and $\kappa_{e,2}$ are both almost an order of magnitude larger than $\kappa_1$, and, therefore, are not affected by photon numbers up to $n_o \approx 10^4$.

The results in Figure 4C show that the intrinsic optical decay rate $\kappa_{i,1}$ increases with the number of intracavity photons $n_o$ in all nc-Si samples as a consequence of losses by carriers generated by the absorption of light stored in the 2$^{nd}$ mode, as will be described below. The results measured from OMS3 and OMS4 show single linear behavior with a similar slope, which can be modelled by:

$$\kappa_{i,1}(n_o) = \kappa_{i,1}(0) + \frac{\partial \kappa_{i,1}}{\partial n_o} n_o \qquad (1),$$

where, according to Ref. [38]:

$$\frac{\partial \kappa_{i,1}}{\partial n_o} n_o = \sigma_{CA}\left(\frac{c}{n_g}\right) N, \qquad (2)$$

$\sigma_{CA}$ and $n_g$ are the carrier absorption cross section, including absorption to and from the traps and free-carrier absorption, and the group velocity index, respectively. A value of $\partial\kappa_{i,1}/\partial n_o \approx 2 \times 10^7$ s$^{-1}$ photon$^{-1}$ can be extracted from the slope of the linear fit of the data. These results indicate that the carriers are generated in both samples by single-photon absorption involving mid-gap states occupied with similar carrier concentration $N_o$, which in turn is determined by the thermal statistical distribution, the average cross section ($\sigma_o$) and the inter-band recombination rate ($\Gamma_C$). Thus, it is possible to express the optically excited carrier density in nc-Si $N$ as:

$$N \approx \left(\frac{\sigma_o N_o}{\Gamma_C}\right) n_o \qquad (3)$$

The c-Si reference sample does not show a significant dependence of $\kappa_{i,1}$ on $n_o$. The effects of trap-assisted carrier generation and two-photon absorption on $\kappa_{i,1}$ seem to be negligible and is beyond the detection limit of our experiment. We can neglect the effect of TPA in the nc-Si OMCs, since the probability of this absorption mechanism is very low and the impact would be comparable to c-Si. The results show that the concentration of gap states in the c-Si OMC is much lower than in the nc-Si OMCs, which is expected. In the sample OMS1, two linear regimes appear and the transition between them occurs at rather low number of intra-cavity photons of $n_{o,t}$=0.2x10$^3$, with the weaker dependence of $\partial\kappa_{i,1}/\partial n_o$ occurring at higher $n_o$. The slope of the linear fit in the region with high photon number is similar to the other nc-Si samples, i.e., $\partial\kappa_{i,1}/\partial n_o \approx$2x10$^7$ s$^{-1}$ photon$^{-1}$. Therefore, in all the nc-Si samples the slope suggests that the gap states in the grain boundaries have with similar concentration, average cross section, decay rate and occupation probability. It is interesting to note that, while the linear slope of the optical power dependent losses seems to be independent of $T_a$, the offset value is seen to decrease with increasing annealing temperature. Thus, scattering at the boundaries or direct photon absorption involving the intra-gap interface states may be the dominating factors defining the behavior of the losses at $n_o$=0 in the nc-Si OMCs, i.e., in the absence of photons in the cavity. [39] The dependence of the intrinsic losses on the photon population also shows that the number of thermally populated trap states is relatively large and that they refill thermally in a short time scale because no saturation effects can be seen in the measurements. Nevertheless, the microscopic mechanisms involving the dependence of the trap capture cross-section and the trap density on the annealing temperature remain to be investigated. The steep slope observed in OMS1 at low $n_o$ seems to be related to a different kind of defect state with much larger cross section, and with depletion of the carrier population when the number of photons in the cavity approaches to a saturation value denoted by $n_o \approx n_{o,t}$.

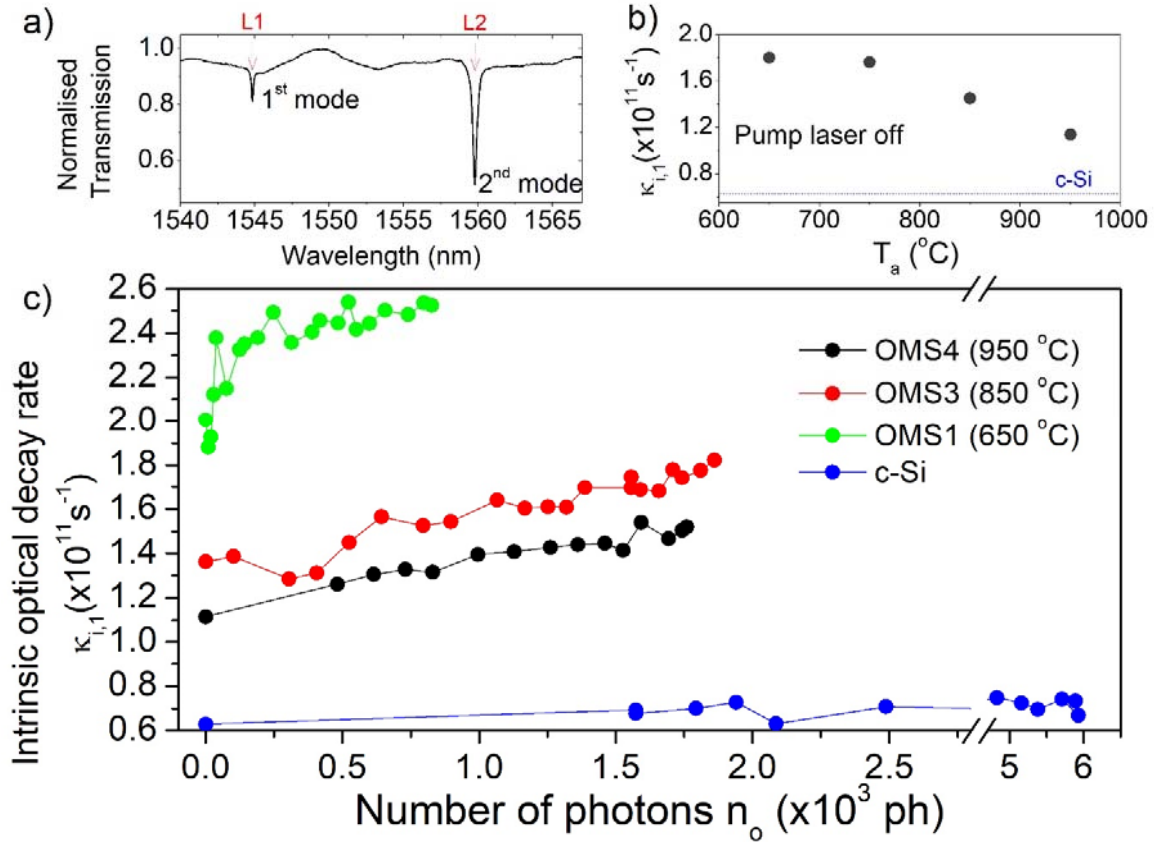

**Figure 4. Optical transmission and intrinsic optical decay rate of the 1st order mode**. (A) Typical optical transmission spectrum of OMC (OMS4) under study displaying the first two optical modes excited with lasers L1 and L2, respectively, in a pump-and-probe experiment to determine the intrinsic optical decay rate $\kappa_{i,1}$. (B) Intrinsic optical loss rate when the pump laser L2 is switched off, i.e., the second mode photon population is set to $n_o=0$, as a function of the annealing temperature of the nc-Si layers. The dashed horizontal line corresponds to the value measured in the c-Si reference sample. (C) Intrinsic optical loss rate of the fundamental mode as a function of the intra-cavity photon population.

Considering that the heat dissipation has a direct impact on the frequency bandwidth of the self-pulsing dynamics occurring in the OMCs, [19] the behavior of heat dissipation rates of the OMCs were studied as a function of $T_a$. A single laser configuration was used to excite the fundamental optical mode $\lambda_{r,1}$ and adjusted $P_{in}$ to obtain $n_{o,max} \approx 2700$ in all samples. As expected, a red-shift of $\lambda_{r,1}$ associated to the thermo-optical (TO) dispersion was observed in response to an effective temperature increase $\Delta T$ in the cavity region. The contribution of the TO effect is the origin of the saw-tooth spectral shape of the transmission obtained when measuring a spectrum by sweeping from short to long wavelengths, with the longest wavelength so that $\lambda_{L1}=\lambda_{r,1}$ and $n_o=n_{o,max}$, as shown in the inset to Figure 5. For comparison, the inset also shows the optical transmission spectrum obtained at low $P_{in}$. Under these experimental conditions it is possible to compare the contribution of the TO effect in each OMC by measuring the maximum $\lambda_{r,1}$ shift. Since the TO coefficient was determined in a previous work to be $\partial \lambda_{r,1}/\partial \Delta T = 0.09$ nm/K, regardless of the crystalline arrangement of the nc-Si films, [19] the spectral shifts can be directly related to $\Delta T$ (see Table I). At this point, it is necessary to recall the results of Figure 4C, which show that the dependence of the optical loss rate on $n_o$ is similar in all the nc-Si OMCs, i.e., that the carrier concentration does not depend on $T_a$. An exception is the high slope region observed in OMS1 at low $n_o$, the contribution of which to the carrier concentration can be neglected given that the number of photons exceeds by an order of magnitude the saturation value $n_{o,t}$.

We have previously demonstrated that the free-carrier-absorption and the subsequent carrier relaxation within the conduction band is the main heat source in single crystal Si OMCs. [4,19,20] In nc-Si, the density of the interface states can easily be above $10^{20}$ eV$^{-1}$cm$^{-3}$, which has an important role in increasing inter-band carrier recombination $\Gamma_c$ and the material heating rates and, consequently, the relative importance of free-carrier absorption as a heating mechanism is reduced. Thus, by considering that both heating processes should be considered in nc-Si, we can deduce that, at equilibrium the temperature increase in nanobeam OMCs is given by:

$$\Delta T \approx \frac{1}{\Gamma_{th}}(\sigma_o N_o n_o)\left(\frac{\alpha_{CA} n_o}{\Gamma_C}+\beta_C\right) \qquad (4)$$

, where $\alpha_{CA}$ is defined as the rate of temperature increase per photon and unit of optically excited carrier density associated to CA, including both the trap-assisted and free-carrier absorption, $\beta_C$ as the rate of temperature increase per unit of optically excited carrier density associated to interband recombination and $\Gamma_{th}$ as the rate at which heat leaks out of the cavity, either through the anchor points at the ends of the OMC or by convection and or conduction to the surrounding environment. In Figure 5, we plot $\Delta T$ for a fixed value of $n_o$ as a function of $T_a$. Since, based on the results shown in Figure 4C, it can be inferred that $\sigma_o N_o / \Gamma_C$ has a weak dependence on $T_a$, and assuming that $\alpha_{CA}$ and $\beta_C$ are also independent of $T_a$, the differences in $\Delta T$ can be associated to variations in $\Gamma_{th}$. Therefore, there is a clear increase of $\Gamma_{th}$ with $T_a$, which is directly connected to the increase of the thermal diffusivity and conductivity of the OMC. The increase in nanocrystallite size with $T_a$ is consistent with the measured increase in thermal conductance of the OMC since smaller nanocrystallites lead to enhanced phonon scattering at the grain boundaries and hence to lower thermal conductivity. [40–42] Therefore, samples with smaller nanocrystallites, i.e., corresponding to lower $T_a$ in this case, have lower thermal conductivity and conductance. A direct comparison of $\Gamma_{th}$ between the nc-Si and c-Si OMCs is not possible because Equation (4) does not hold for single crystal silicon since it implies that carriers are generated by single photon absorption. Nevertheless, the change in temperature $\Delta T$ extracted for c-Si is much smaller than those of nc-Si and it is deduced that $\Gamma_{th}$ is much smaller in nc-Si as a consequence of phonon scattering at the grain boundaries. [43] A dependence of $\Gamma_{th}$ on $T_a$ consistent with data shown in Figure 5 was detected independently by measuring the time-domain thermo-reflectivity of nc-Si membranes (see Supplementary Information).

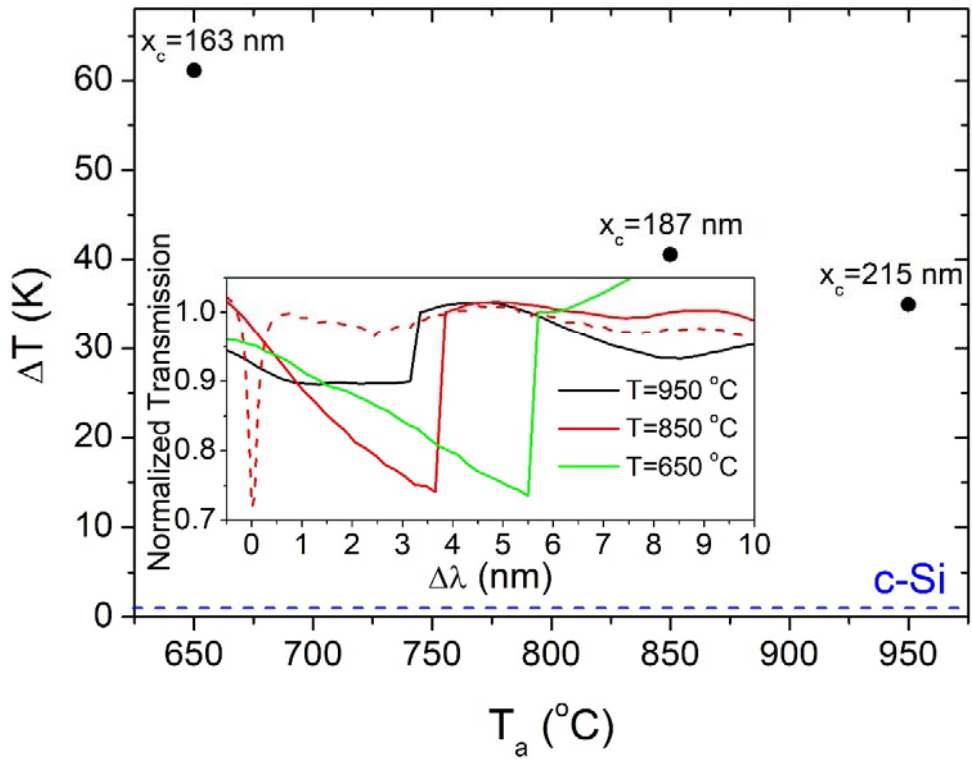

**Figure 5. Temperature increase of OMC nanobeams fabricated in nc-SOI wafers with different annealing temperature $T_a$ for a fixed photon number of $2.7 \times 10^3$.** The dashed horizontal line corresponds to the value extracted from the measurements of the c-Si reference sample. The average nanocrystallite size ($x_c$) is also shown. The inset shows the transmission curves with respect to the resonance value, at low laser power, obtained at the same value of the maximum intra-cavity photon number ($n_{o,max}$). The transmission signals, which arise from the interplay of the FCA and the TO effect, are normalized to the values just after the resonance is lost. The red dashed curve illustrates a transmission curve at low laser power measured from sample OMS3 ($T_a$=850 ºC) so that no significant TO effect is observed.

4. **Conclusions**

In summary, optical, mechanical and thermal properties and, especially, the dissipation rates of optomechanical nanobeams fabricated in nanocrystalline silicon - on - insulator wafers have been investigated. The annealing temperature of the originally amorphous silicon films is found to have a strong effect on the material properties and, consequently, on the behavior of the optomechanical cavities. The structural properties were characterized by transmission electron microscopy, picosecond acoustic methods and Raman spectroscopy. The films consisted of randomly oriented crystallites ranging from a few nm to a few hundreds of nm. The size distribution shifts towards larger crystalline domains with increasing annealing temperature, which yielded a smaller volume fraction of grain boundaries within the layers. This appears to be the origin of the extracted tendencies of optical, mechanical and thermal decay rates, suggesting that the dominant factors are associated with physical phenomena occurring within and at the grain boundaries and are related to the mid-gap states. Regarding the optical losses of the photonic cavities, it is concluded that intrinsic damping losses decrease with annealing temperature, probably due to a reduced scattering at grain boundaries and to the effect of annealing on mid-gap interface states. The best optical and mechanical quality factors at room temperature, $1.3 \times 10^4$ at 1.5 μm wavelength (around 200 THz) and $1.7 \times 10^3$ at 2.4 GHz, were found in OMCs fabricated on nc-Si wafers annealed at 950 ºC, i.e., with the smallest grain boundary volume. The relatively strong carrier absorption losses are due to carriers generated by single-photon absorption mediated by the mid-gap traps at the grain boundaries, which are independent of the annealing temperature in the studied range. The thermal decay rate significantly increases with the annealing temperature due to enhanced thermal conductivity and diffusivity, which can be associated to phonon scattering at the grain boundaries.

The mechanical decay rates seem to be dominated by thermoelastic dissipation, which depends directly on the thermal diffusivity and, therefore, decreases with annealing temperature.

In a comparison to results from cavities made on SOI wafers, i.e., of single crystal silicon, of identical geometry, it was established that the main differences were a more efficient optical absorption by single photons at the telecom wavelength of 1.5 μm, a stronger recombination via mid-gap states and lower thermal conductivity, all leading to an enhanced non-linear behaviour, especially at room temperature.

Finally, this work lends support to technological approaches using nc-Si since the layer structure and its thickness can be tuned to optimise desired device properties. In particular, these results indicate that engineering the grain or crystallite size- and grain boundary-dependent properties provides a powerful way to tune the properties, and therefore the performance, of opto-mechanical systems.

**Associated Content**

Further details on the analysis of the dark field TEM images and of the TDTR measurements are provided in Supporting Information.

**Author Information**

Corresponding Author

*E-mail: dnavarro@ub.edu**Funding Sources**

The following support is gratefully acknowledged: the European Commission project PHENOMEN (H2020-EU- FET Open GA no. 713450), the Spanish Severo Ochoa Excellence program (SEV-2017-0706), CMST and ECA: the Spanish MICINN project SIP (PGC2018-101743-B-I00), DNU and AM: the Spanish MICINN project PGC2018-094490-B-C22. DNU holds a Ramón y Cajal postdoctoral fellowship (RYC-2014-15392), MFC and GA hold a S. Ochoa and a M. S. Curie COFUND BIST postgraduate studentship, respectively.

**Acknowledgements**

The authors thank F. J. Belarre Triviño and B. Ballesteros Pérez for the TEM sample preparation and images.

**References**

[1]  M. Aspelmeyer, T. J. Kippenberg, and F. Marquardt, *Cavity Optomechanics*, Rev. Mod. Phys. **86**, 1391 (2014).

[2]  Y. T. Yang, C. Callegari, X. L. Feng, K. L. Ekinci, and M. L. Roukes, *Zeptogram-Scale Nanomechanical Mass Sensing*, Nano Lett. **6**, 583 (2006).

[3]  M. Bagheri, M. Poot, M. Li, W. P. H. Pernice, and H. X. Tang, *Dynamic Manipulation of Nanomechanical Resonators in the High-Amplitude Regime and Non-Volatile Mechanical Memory Operation*, Nat. Nanotechnol. **6**, 726 (2011).

[4]  D. Navarro-Urrios, N. E. Capuj, M. F. Colombano, P. D. Garciá, M. Sledzinska, F. Alzina, A. Griol, A. Martínez, and C. M. Sotomayor-Torres, *Nonlinear Dynamics and Chaos in an Optomechanical Beam*, Nat. Commun. **8**, (2017).

[5]  M. F. Colombano, G. Arregui, N. E. Capuj, A. Pitanti, J. Maire, A. Griol, B. Garrido, A. Martinez, C. M. Sotomayor-Torres, and D. Navarro-Urrios, *Synchronization of*

*Optomechanical Cavities by Mechanical Interaction*, (2018).

[6] G. Heinrich, M. Ludwig, J. Qian, B. Kubala, and F. Marquardt, *Collective Dynamics in Optomechanical Arrays*, Phys. Rev. Lett. **107**, 043603 (2011).

[7] M. Davanço, S. Ates, Y. Liu, and K. Srinivasan, *Si3N4 Optomechanical Crystals in the Resolved-Sideband Regime*, Appl. Phys. Lett. **104**, 41101 (2014).

[8] K. C. Balram, M. I. Davanço, J. D. Song, and K. Srinivasan, *Coherent Coupling between Radiofrequency, Optical and Acoustic Waves in Piezo-Optomechanical Circuits*, Nat. Photonics **10**, 346 (2016).

[9] J. Bochmann, A. Vainsencher, D. D. Awschalom, and A. N. Cleland, *Nanomechanical Coupling between Microwave and Optical Photons*, Nat. Phys. **9**, 712 (2013).

[10] C. Xiong, W. Pernice, X. Sun, C. Schuck, K. Y. Fong, and H. Tang, *Aluminum Nitride as a New Material for Chip-Scale Optomechanics and Nonlinear Optics*, Vol. 14 (2012).

[11] M. J. Burek, J. D. Cohen, S. M. Meenehan, N. El-Sawah, C. Chia, T. Ruelle, S. Meesala, J. Rochman, H. A. Atikian, M. Markham, D. J. Twitchen, M. D. Lukin, O. Painter, and M. Lončar, *Diamond Optomechanical Crystals*, Optica **3**, 1404 (2016).

[12] M. Mitchell, B. Khanaliloo, D. P. Lake, T. Masuda, J. P. Hadden, and P. E. Barclay, *Single-Crystal Diamond Low-Dissipation Cavity Optomechanics*, Optica **3**, 963 (2016).

[13] M. Eichenfield, J. Chan, R. M. Camacho, K. J. Vahala, and O. Painter, *Optomechanical Crystals*, Nature **462**, 78 (2009).

[14] D. Navarro-Urrios, J. Gomis-Bresco, S. El-Jallal, M. Oudich, A. Pitanti, N. Capuj, A. Tredicucci, F. Alzina, A. Griol, Y. Pennec, B. Djafari-Rouhani, A. Martínez, and C. M. Sotomayor Torres, *Dynamical Back-Action at 5.5 GHz in a Corrugated Optomechanical Beam*, AIP Adv. **4**, (2014).

[15] J. Chan, T. P. M. Alegre, A. H. Safavi-Naeini, J. T. Hill, A. Krause, S. Gröblacher, M. Aspelmeyer, and O. Painter, *Laser Cooling of a Nanomechanical Oscillator into Its Quantum Ground State*, Nature **478**, 89 (2011).

[16] R. Riedinger, S. Hong, R. A. Norte, J. A. Slater, J. Shang, A. G. Krause, V. Anant, M. Aspelmeyer, and S. Gröblacher, *Non-Classical Correlations between Single Photons and Phonons from a Mechanical Oscillator*, Nature **530**, 313 (2016).

[17] G. Harbeke, *Growth and Physical Properties of LPCVD Polycrystalline Silicon Films*, J. Electrochem. Soc. **131**, 675 (1984).

[18] M. Ylönen, A. Torkkeli, and H. Kattelus, *In Situ Boron-Doped LPCVD Polysilicon with Low Tensile Stress for MEMS Applications*, Sensors Actuators A Phys. **109**, 79 (2003).

[19] D. Navarro-Urrios, N. E. Capuj, J. Maire, M. Colombano, J. Jaramillo-Fernandez, E. Chavez-Angel, L. L. Martin, L. Mercadé, A. Griol, A. Martínez, C. M. Sotomayor-Torres, and J. Ahopelto, *Nanocrystalline Silicon Optomechanical Cavities*, Opt. Express **26**, (2018).

[20] D. Navarro-Urrios, N. E. Capuj, J. Gomis-Bresco, F. Alzina, A. Pitanti, A. Griol, A. Martínez, and C. M. Sotomayor Torres, *A Self-Stabilized Coherent Phonon Source Driven by Optical Forces*, Sci. Rep. **5**, (2015).

[21] G. G. Stoney and C. A. Parsons, *The Tension of Metallic Films Deposited by Electrolysis*, Proc. R. Soc. London. Ser. A, Contain. Pap. a Math. Phys. Character **82**, 172 (1909).

[22] G. C. A. M. Janssen, M. M. Abdalla, F. van Keulen, B. R. Pujada, and B. van Venrooy, *Celebrating the 100th Anniversary of the Stoney Equation for Film Stress: Developments from Polycrystalline Steel Strips to Single Crystal Silicon Wafers*, Thin Solid Films **517**, 1858 (2009).

[23] D. E. Aspnes and A. A. Studna, *Dielectric Functions and Optical Parameters of Si, Ge, GaP, GaAs, GaSb, InP, InAs, and InSb from 1.5 to 6.0 EV*, Phys. Rev. B **27**, 985 (1983).

[24] E. Iwase, P.-C. Hui, D. Woolf, A. W. Rodriguez, S. G. Johnson, F. Capasso, and M. Lončar, *Control of Buckling in Large Micromembranes Using Engineered Support Structures*, J. Micromechanics Microengineering **22**, 065028 (2012).

[25] I. Theodorakos, I. Zergioti, V. Vamvakas, D. Tsoukalas, and Y. S. Raptis, *Picosecond and Nanosecond Laser Annealing and Simulation of Amorphous Silicon Thin Films for Solar Cell Applications*, J. Appl. Phys. **115**, 43108 (2014).

[26] I. H. Campbell and P. M. Fauchet, *The Effects of Microcrystal Size and Shape on the One Phonon Raman Spectra of Crystalline Semiconductors*, Solid State Commun. **58**, 739 (1986).

[27] S. Veprek, F.-A. Sarott, and Z. Iqbal, *Effect of Grain Boundaries on the Raman Spectra, Optical Absorption, and Elastic Light Scattering in Nanometer-Sized Crystalline Silicon*, Phys. Rev. B **36**, 3344 (1987).

[28] H. Richter, Z. P. Wang, and L. Ley, *The One Phonon Raman Spectrum in Microcrystalline*

*Silicon*, Solid State Commun. **39**, 625 (1981).

[29] P. A. Mante, J. F. Robillard, and A. Devos, *Complete Thin Film Mechanical Characterization Using Picosecond Ultrasonics and Nanostructured Transducers: Experimental Demonstration on SiO2*, Appl. Phys. Lett. **93**, 71909 (2008).

[30] J. Gomis-Bresco, D. Navarro-Urrios, M. Oudich, S. El-Jallal, A. Griol, D. Puerto, E. Chavez, Y. Pennec, B. Djafari-Rouhani, F. Alzina, A. Martínez, and C. M. S. Torres, *A One-Dimensional Optomechanical Crystal with a Complete Phononic Band Gap*, Nat. Commun. **5**, (2014).

[31] S. S. Verbridge, J. M. Parpia, R. B. Reichenbach, L. M. Bellan, and H. G. Craighead, *High Quality Factor Resonance at Room Temperature with Nanostrings under High Tensile Stress*, J. Appl. Phys. **99**, 124304 (2006).

[32] Y. Sun, D. Fang, and A. K. Soh, *Thermoelastic Damping in Micro-Beam Resonators*, Int. J. Solids Struct. **43**, 3213 (2006).

[33] C. M. Zener and S. Siegel, *Elasticity and Anelasticity of Metals.*, J. Phys. Colloid Chem. **53**, 1468 (1949).

[34] V. T. Srikar and S. D. Senturia, *Thermoelastic Damping in Fine-Grained Polysilicon Flexural Beam Resonators*, J. Microelectromechanical Syst. **11**, 499 (2002).

[35] S. S. Verbridge, D. F. Shapiro, H. G. Craighead, and J. M. Parpia, *Macroscopic Tuning of Nanomechanics: Substrate Bending for Reversible Control of Frequency and Quality Factor of Nanostring Resonators*, Nano Lett. **7**, 1728 (2007).

[36] S. Kumar and M. Aman Haque, *Stress-Dependent Thermal Relaxation Effects in Micro-Mechanical Resonators*, Acta Mech. **212**, 83 (2010).

[37] D. Macdonald and A. Cuevas, *Validity of Simplified Shockley-Read-Hall Statistics for Modeling Carrier Lifetimes in Crystalline Silicon*, Phys. Rev. B **67**, 75203 (2003).

[38] T. J. Johnson, M. Borselli, and O. Painter, *Self-Induced Optical Modulation of the Transmission through a High-Q Silicon Microdisk Resonator*, Opt. Express **14**, 817 (2006).

[39] H. C. Card, *The Photoconductivity of Polycrystalline Semiconductors*, J. Appl. Phys. **52**, 3671 (1981).

[40] S. Uma, A. D. McConnell, M. Asheghi, K. Kurabayashi, and K. E. Goodson, *Temperature-Dependent Thermal Conductivity of Undoped Polycrystalline Silicon Layers*, Int. J. Thermophys. **22**, 605 (2001).

[41] H. Dong, B. Wen, and R. Melnik, *Relative Importance of Grain Boundaries and Size Effects in Thermal Conductivity of Nanocrystalline Materials*, Sci. Rep. **4**, 7037 (2014).

[42] B. Jugdersuren, B. T. Kearney, D. R. Queen, T. H. Metcalf, J. C. Culbertson, C. N. Chervin, R. M. Stroud, W. Nemeth, Q. Wang, and X. Liu, *Thermal Conductivity of Amorphous and Nanocrystalline Silicon Films Prepared by Hot-Wire Chemical-Vapor Deposition*, Phys. Rev. B **96**, 1 (2017).

[43] M. Nomura, Y. Kage, J. Nakagawa, T. Hori, J. Maire, J. Shiomi, R. Anufriev, D. Moser, and O. Paul, *Impeded Thermal Transport in Si Multiscale Hierarchical Architectures with Phononic Crystal Nanostructures*, Phys. Rev. B **91**, 205422 (2015).